\documentclass[11pt]{article}

\usepackage[margin=0.82in]{geometry}
\usepackage[T1]{fontenc}
\usepackage[utf8]{inputenc}
\usepackage{lmodern}
\usepackage{microtype}
\usepackage{amsmath,amssymb,mathtools}
\usepackage{booktabs,longtable,tabularx,array,multirow}
\usepackage{graphicx}
\usepackage{enumitem}
\usepackage{xcolor}
\usepackage[normalem]{ulem}
\usepackage{hyperref}
\usepackage{url}
\usepackage{float}
\usepackage{caption}
\usepackage{subcaption}
\usepackage{fancyhdr}
\usepackage{titlesec}
\usepackage{listings}
\usepackage{ragged2e}

\definecolor{navy}{HTML}{17324D}
\definecolor{blue}{HTML}{2D5B89}
\definecolor{teal}{HTML}{2E6F73}
\definecolor{softgray}{HTML}{F3F5F7}
\definecolor{midgray}{HTML}{5C6670}
\definecolor{darkred}{HTML}{8C2F39}
\definecolor{darkgreen}{HTML}{2F6B4F}
\definecolor{editred}{HTML}{C00000}

\hypersetup{
  colorlinks=true,
  linkcolor=navy,
  citecolor=teal,
  urlcolor=blue,
  pdftitle={BeTaL-GBI: Admission-Aware Benchmark Tuning and Full-Stack Verification of Geometric Belief Interfaces},
  pdfauthor={Alvin Spivey and Yu Huang}
}

\titleformat{\section}{\Large\bfseries\color{navy}}{\thesection}{0.6em}{}
\titleformat{\subsection}{\large\bfseries\color{blue}}{\thesubsection}{0.6em}{}
\titleformat{\subsubsection}{\normalsize\bfseries\color{teal}}{\thesubsubsection}{0.6em}{}
\setlist[itemize]{leftmargin=1.4em,itemsep=2pt,topsep=4pt}
\setlist[enumerate]{leftmargin=1.6em,itemsep=2pt,topsep=4pt}
\newcommand{\BeTaL}{\textsc{BeTaL}}

\newcommand{\EV}{\mathrm{EV}}
\newcommand{\rhatadm}{\widehat{\rho}_{\mathrm{adm}}}
\newcommand{\rhattask}{\widehat{\rho}_{\mathrm{task}}}

\newcommand{\code}[1]{\texttt{#1}}

\newenvironment{claimbox}{%
  \begin{center}\begin{minipage}{0.94\linewidth}\color{navy}\hrule\vspace{0.6em}\color{black}\small
}{%
  \vspace{0.5em}\color{navy}\hrule\end{minipage}\end{center}
}

\title{\textbf{BeTaL-GBI: Admission-Aware Benchmark Tuning and Full-Stack Verification of Geometric Belief Interfaces}\\[0.35em]
\large A Companion Empirical Study of BoundaryBench v0.2, GBI v2, and GBI-DCSE v3}
\author{Alvin Spivey \& Yu Huang\\Light Imaging Technologies, Inc.\\\texttt{alvin@lightimagingtech.com}}
\date{August 20, 2026}

\begin{document}
\maketitle

\begin{abstract}
A verification substrate is more credible when it can expose errors in claims about itself, not only errors in model output. In extending GBI-DCSE into executable claim-level verification, the v3 harness falsified a numerical implication in the original architecture manuscript: the previously reported Fisher-conditioning value \(\epsilon\approx0.066\) satisfies the declared \(\kappa_2\le10^4\) budget only on the slice \([\epsilon,3,4,5]\), whereas the full declared box \([\epsilon,20]^4\) requires \(\epsilon\approx0.326472\). The mismatch is preserved as an \texttt{ERRATUM} rather than silently retuned. This self-correction motivates the broader question of the paper: can an enterprise verification architecture distinguish interface failure, task competence, policy admissibility, and control-plane integrity while making unsupported claims auditable?

The frozen BoundaryBench v0.1 experiment supplied the starting failure mode. Qwen3-4B-Instruct-2507 completed 768 held-out executions, but none crossed the structured contract: 369 failed safe parsing and 399 failed schema validation, so every execution was quarantined. That establishes fail-closed handling for one frozen model/configuration, not general protection from hallucination; zero admission also makes downstream selectivity and task difficulty unidentifiable.

This companion study evaluates three successive refinements. First, \BeTaL-GBI v0.2 adapts Benchmark Tuning with an LLM-in-the-loop (BeTaL) to a versioned enterprise admission boundary. It parameterizes nine benchmark dimensions over a finite design space of 2,218,750,380 grid points and separates format admission from conditional task performance: \(\rhatadm=N_{\mathrm{admitted}}/N\) and \(\rhattask=N_{\mathrm{verified}}/N_{\mathrm{admitted}}\). Task difficulty is deliberately undefined below a 0.05 admission floor. After a declared schema-compliant repair restores admission, a model-free feedback search reaches a mean held-out target gap of 2.87\% across four target levels, versus 13.61\% and 11.46\% for two reimplemented non-feedback baselines. No language model is executed in this phase.

Second, GBI v2 replaces answer-key agreement as the runtime admission predicate with a reference-independent witness state \(W\) and executable policy \(P\). On 512 synthetic tasks, the declared 16-gate policy at strictness 0.6, selected by a predeclared rule, detects all 116 injected severe contradictions and refuses none of 99 clean admissible records under the narrow predeclared denominator; a broader denominator yields 5/117 = 4.27\%, and both are retained. A confident-hallucinator surrogate makes 221 attempted over-admissions with zero silent promotions, while an evidence-forger surrogate is refused on all 512 cases. Deterministic format repair raises admission from 0 to 46.48\% without raising verified completion above zero, separating syntactic recovery from substantive correctness.

Third, GBI-DCSE v3 maps 99 manuscript claims to machine-readable evidence classes and resolvers. Ninety-five of 96 testable registered claims meet their declared acceptance criteria, three claims remain explicitly out of scope, and the Fisher statement is retained as one erratum. The standalone verifier executes 148 declared checks with zero implementation failures; a check can pass by correctly reproducing a value, detecting a fault, or confirming an \texttt{ERRATUM}/\texttt{OUT OF SCOPE} classification, so 148/148 passing checks is not a claim that all 99 registered assertions are true. The harness also exercises signed-ledger behavior, attestation failure classes, a PBFT-style quorum model, sparse certificate forgery tests, atomic rollback, replayable receipts, and a second synthetic domain.

The resulting claim is deliberately bounded. Under the declared synthetic populations, GBI-DCSE behaves as a selective, policy-versioned, self-auditing test and routing substrate. BoundaryBench v0.2, GBI v2, and DCSE v3 execute no new language model and do not establish clinical safety, production readiness, correctness of institutional policy, or robustness to arbitrary adaptive attackers. Those system boundaries are stated once near the beginning and treated as part of the evaluation contract.
\end{abstract}

\tableofcontents
\newpage

\section{From fail-closed containment to self-auditing verification}
\label{sec:introduction}

The architecture manuscript introduced the Geometric Belief Interface (GBI) and Decentralized Cryptographic Sheaf-Enclave (DCSE) as a boundary between untrusted discovery and governed enterprise action \cite{spivey2026gbi}. Its central design rule is that a model output is \emph{evidence, not authority}. A model, parser, or agent may propose a typed local state; deterministic checks over identity, provenance, terminology or schema version, temporal validity, dependencies, evidence, and policy determine what may proceed.

The central methodological lesson of the companion study is stricter: the same discipline must apply to claims made by the verification system and its authors. A system that catches malformed model output but silently preserves a falsified numerical assumption has not completed the verification loop.

\subsection{A verifier that falsified one of its own numerical claims}

The strongest illustration emerged from the Fisher-conditioning check. The architecture paper reported \(\epsilon\approx0.066\) near the condition-number budget \(\kappa_2\le10^4\). V3 reproduced that value for the stated one-dimensional slice
\[
\alpha=(\epsilon,3,4,5),
\]
but then evaluated the stronger box-wide criterion that the manuscript itself prescribed. On \([\epsilon,20]^4\), the slice value permits a worst-corner condition number of approximately \(4.79\times10^5\), about 48 times the declared budget. Solving the declared box-wide criterion yields
\[
\epsilon_{\mathrm{box}}\approx0.326472.
\]
Rather than modifying the acceptance criterion after seeing the result, the claim register marks the original implication as \texttt{ERRATUM}, preserves the reproduced slice value, and records the corrected box-wide bound. Section~\ref{sec:fisher-erratum} gives the full calculation.

\begin{claimbox}
\textbf{Self-correction criterion.} A verification substrate should make its own failed assertions visible and attributable. In this study, the erratum is therefore a positive result about the \emph{claim-to-evidence process}, even though the underlying numerical assertion is a failed claim.
\end{claimbox}

\subsection{Operational stakes in the healthcare instantiation}

The mathematical boundary becomes easier to interpret when tied to the type of governed state change it is meant to precede. Table~\ref{tab:clinical-cases} gives two \emph{synthetic policy-routing examples}. They are not treatment recommendations and the policies are illustrative; their purpose is to show what it means for \(W\) and \(P\) to control an EHR write rather than allowing a model score to act as authority.

\begin{table}[H]
\centering
\small
\caption{Two synthetic healthcare boundary cases used to ground the abstract verification objects. Clinical facts cited here motivate the evidence fields; the GBI policy examples remain institution-specific and non-clinical.}
\label{tab:clinical-cases}
\begin{tabularx}{\linewidth}{>{\RaggedRight\arraybackslash}p{0.18\linewidth} >{\RaggedRight\arraybackslash}p{0.27\linewidth} >{\RaggedRight\arraybackslash}p{0.28\linewidth} >{\RaggedRight\arraybackslash}X}
\toprule
Synthetic proposal & Authoritative context & Example policy action & What GBI does not decide\\
\midrule
Amoxicillin medication request with a record of serious penicillin/beta-lactam hypersensitivity & FHIR \code{AllergyIntolerance} is designed to represent patient-specific risk from substance exposure and can inform clinical decision support; current U.S. AMOXIL labeling lists a history of serious hypersensitivity to amoxicillin or other beta-lactams as a contraindication \cite{fhirAllergy,fdaAmoxil}. & A version-pinned local policy could refuse autonomous commit and require named clinical review or an authorized override with provenance. & It does not diagnose an allergy, select an alternative antibiotic, or determine patient-specific treatment.\\
\addlinespace
Metformin medication request with required renal-function evidence absent or stale & Current DailyMed labeling directs renal-function assessment before initiation and periodically thereafter, with explicit eGFR-related restrictions \cite{dailymedMetformin}. & If the institution's declared \(P\) requires qualifying renal context, missing evidence can route the proposed write to abstention, quarantine, or expert review until the required evidence is present. & It does not set an individualized dose, diagnose renal impairment, or independently decide whether metformin is clinically appropriate.\\
\bottomrule
\end{tabularx}
\end{table}

These cases clarify the role of the geometry and cryptography later in the paper. The discovery system may emit a proposal; the verification substrate determines whether the proposal has the identity binding, evidence, version, provenance, dependency state, and authority required by the institution's policy before an atomic transaction is constructed. Geometry can diagnose inconsistency and cryptography can bind/replay the decision, but neither substitutes for clinical judgment.

\subsection{Questions left by the frozen v0.1 experiment}

BoundaryBench v0.1 translated the architecture into a frozen empirical interface. The held-out model host saw answer-key-free inputs, while a trusted verifier scored frozen outputs after execution. All 768 Qwen3-4B-Instruct-2507 executions completed, but zero model results satisfied the admissibility contract: 369 were safe parse rejects and 399 were safe schema rejects. The result shows that the observed malformed/schema-invalid proposals did not silently become downstream actions. It does not establish containment of arbitrary hallucinations, and zero coverage does not establish low selective risk.

The zero-coverage result creates three scientific questions that v0.1 cannot answer:
\begin{enumerate}
  \item \textbf{Can benchmark difficulty be tuned if the model never reaches the task-semantic layer?} A target-performance optimizer has no useful signal if every output dies at formatting.
  \item \textbf{Can the GBI gate be selective rather than merely fail closed?} A system that refuses everything catches every bad item vacuously but has no operational utility.
  \item \textbf{Do the broader DCSE claims survive executable scrutiny?} The original paper contains claims about ledger non-equivocation, attestation, BFT fallback, sparse enclave checks, transaction atomicity, receipts, and portability that v0.1 did not test.
\end{enumerate}

\subsection{System boundaries and non-goals}
\label{sec:boundaries}

Table~\ref{tab:system-boundaries} defines the evidentiary perimeter once. Later sections retain only local qualifiers that are necessary to interpret a particular result.

\begin{table}[H]
\centering
\small
\caption{System boundaries for the companion evaluation.}
\label{tab:system-boundaries}
\begin{tabularx}{\linewidth}{>{\bfseries\RaggedRight\arraybackslash}p{0.18\linewidth} >{\RaggedRight\arraybackslash}p{0.37\linewidth} >{\RaggedRight\arraybackslash}X}
\toprule
Boundary & Evaluated here & Not claimed\\
\midrule
Model behavior & Frozen v0.1 Qwen outputs are a historical empirical baseline; v0.2--v3 evaluate benchmark and verification machinery. & General hallucination containment, frontier-model capability, or new LLM performance in v0.2--v3.\\
Clinical authority & Synthetic witness/policy states route proposed EHR actions and expose the reason for refusal/review. & Diagnosis, treatment recommendation, autonomous prescribing, clinical safety, medical-device validation, or replacement of clinician judgment.\\
Policy authority & Deterministic execution, version binding, ablations, replay, and reference independence of the declared \(P\). & That an institution's policy is clinically, legally, or ethically correct; a verifier can faithfully execute a bad policy.\\
Infrastructure & Software models/proxies exercise signatures, attestation checks, quorum arithmetic, certificates, rollback, receipts, and FHIR-like transaction semantics. & Live TEE attestation, a running BFT cluster, production FHIR integration, hardware-root security, or a zero-knowledge proof system.\\
Mathematical diagnostics & Finite Boolean, Fisher, mapping-cone, and certificate properties under declared numerical tests. & Clinical truth from geometry; mapping-cone diagnostics have zero scoring weight.\\
Portability & Code-level reuse is measured in a second synthetic sensitive-infrastructure instantiation. & Readiness for real healthcare, finance, government, or classified infrastructure without domain-specific authority and validation.\\
Adversarial scope & Declared malformed, over-admitting, evidence-forging, policy-ablation, ledger, attestation, quorum, certificate, and transaction faults. & Completeness against adaptive attackers, side channels, compromised hardware roots, malicious authorities, or unmodeled policy/source-system failures.\\
\bottomrule
\end{tabularx}
\end{table}

\newpage

This companion evaluation addresses those questions in three layers. Figure~\ref{fig:ladder} summarizes the progression.

\begin{table}[H]
\centering
\begin{minipage}{0.92\linewidth}
\color{navy}\hrule\vspace{0.6em}\color{black}
\centering
\textbf{Evaluation ladder}\\[0.7em]
\begin{tabularx}{\linewidth}{>{\bfseries}l >{\RaggedRight\arraybackslash}X}
BoundaryBench v0.1 & Frozen model execution asks whether outputs cross a structured admission contract. Result: 0/768 admitted, exposing a format/schema floor. \\[0.6em]  \vspace{-0.4em} \\
BeTaL-GBI v0.2 & Parameterizes benchmark difficulty, separates admission from task performance, detects degenerate objectives, and measures search/reachability using declared surrogates. \\[0.6em]  \vspace{-0.4em} \\
GBI v2 & Replaces answer-key agreement as the runtime gate with witness-grounded external validity and an executable versioned policy; evaluates selectivity, adversaries, liveness, and mathematical checks. \\[0.6em]  \vspace{-0.4em} \\
GBI-DCSE v3 & Builds a claim register over the broader testable architecture and tests ledger, attestation, consensus, sparse enclave certification, atomic rollback, receipts/replay, and cross-domain portability.
\end{tabularx}
\vspace{0.5em}\color{navy}\hrule\color{black}
\end{minipage}
\caption{The experiments increase the scope of what is evaluated. They do not monotonically increase realism: v0.2--v3 are synthetic and execute no language model.}
\label{fig:ladder}
\end{table}

\begin{claimbox}
\textbf{Primary empirical claim of this companion paper.} Under the declared synthetic populations and executable policy, the later GBI/DCSE evaluations are non-vacuous: they permit eligible work, refuse declared severe contradictions, expose policy ablations, replay decisions from receipts, and classify unsupported or falsified claims explicitly. This is stronger evidence about the \emph{verification substrate} than v0.1, but it is not new evidence about frontier-model capability because no language model is executed in v0.2--v3.
\end{claimbox}

\section{Objects, notation, and evaluation boundary}

\subsection{GBI/DCSE state}

The enterprise state is represented as
\[
(B,V,\Theta,F,W,L,P),
\]
where \(B\) is the finite boundary algebra, \(V\) the versioned semantic bundle, \(\Theta\) the calibrated evidence registry, \(F\) the candidate/local state, \(W\) the authoritative grounding state, \(L\) the identity/provenance/non-equivocation ledger, and \(P\) the runtime admissibility policy \cite{spivey2026gbi}. GBI v2 makes the last two deployment objects particularly explicit: runtime admission depends on authoritative evidence and declared policy, not on the hidden answer key used to score a benchmark after the fact.

A useful operational decomposition is
\[
P=(A,G,T,D,H,Q,R,\Lambda,E,F_b),
\]
with allowed actions \(A\), hard gates \(G\), thresholds and validity windows \(T\), dependencies \(D\), human authority \(H\), quarantine semantics \(Q\), recovery/escalation \(R\), liveness/degraded operation \(\Lambda\), signed exceptions \(E\), and failover \(F_b\). A policy instance is pinned by identifiers such as policy id, version, effective time, authority, and hash.

Table~\ref{tab:clinical-cases} illustrates this object language operationally. In the allergy example, the candidate \(F\) is a proposed medication request, \(W\) includes the signed/versioned allergy and source evidence, \(V\) pins the terminology/schema bundle, and \(P\) specifies whether missing or conflicting evidence requires review rather than commit. In the renal-context example, \(P\) can require a current evidence predicate before the candidate crosses the write boundary. The clinical authority remains outside the mathematical object: \(P\) encodes a policy supplied by an authorized institution; GBI tests and executes that policy rather than inventing it.

\subsection{External validity and benchmark truth are different}

The deployment-facing predicate is modeled as
\[
\EV:B\times W\rightarrow\{0,1\},
\]
with the policy mapping verified state to actions such as admit, historical-only admit, quarantine, abstain, expert review, or reject. This object must be independent of benchmark answer keys if it is to represent a plausible runtime gate.

The benchmark still needs a reference for measurement. The two truths therefore have different roles:

\begin{itemize}
  \item \textbf{Runtime truth surrogate:} witness bundle \(W\) plus versioned policy \(P\), used by the gate.
  \item \textbf{Evaluation reference:} synthetic generator/reference state, used after the gate to quantify agreement or identify stale references after deliberate injection.
\end{itemize}

GBI v2 includes an abstract-syntax-tree probe that finds no forbidden reference access in the gate modules, and a shuffle test that randomly permutes all 512 reference actions while leaving all 512 gate verdicts unchanged. These tests do not prove real-world truth; they show that the executable admission decision is not secretly using the answer key.

\subsection{Scope labels used throughout}

We use the following evidence labels:

\begin{itemize}
  \item \textbf{REPRODUCED}: a published numerical statement is independently recomputed.
  \item \textbf{MEASURED}: an executable deterministic or simulated system property is measured directly in the supplied implementation.
  \item \textbf{MEASURED SYNTH.}: the measured property depends on a declared synthetic population or injected taxonomy.
  \item \textbf{STRUCTURAL}: a code/schema/protocol property is checked structurally rather than by live deployment.
  \item \textbf{PARTIAL PROXY}: a lower-scope stand-in is measured, with the missing deployment mechanism explicit.
  \item \textbf{OUT OF SCOPE}: the required environment or evidence is absent and no proxy is promoted to the target claim.
  \item \textbf{ERRATUM}: an executable assertion contradicts a numerical statement in the architecture manuscript.
\end{itemize}

These labels become first-class in the v3 claim register.

\section{BeTaL-GBI v0.2: benchmark tuning at an admission boundary}

\subsection{Relation to BeTaL}

BeTaL (Benchmark Tuning with an LLM-in-the-loop) parameterizes benchmark design choices and uses an LLM designer to search for benchmark instances with target properties, including difficulty \cite{dsouza2025betal}. The GBI adaptation retains the core environment-design idea but adds an admission-aware objective. This is necessary because an enterprise benchmark may fail before semantic task performance is measurable.

The v0.2 artifact deliberately does \emph{not} execute the BeTaL LLM designer. It provides the designer seam, evaluates a deterministic feedback-coordinate controller, and compares it with two non-feedback search baselines. Consequently, the correct interpretation is ``the parameterized GBI environment produces a usable tuning signal once admission is non-degenerate.''

\subsection{The structural problem: zero admission destroys the difficulty signal}

A naive difficulty objective would tune configuration \(v\in\mathcal V\) toward target performance \(\rho\) by minimizing
\[
g(v)=|\widehat{\rho}(v)-\rho|.
\]
That is ill posed when every output is rejected by a format gate. BeTaL-GBI therefore factors the rate:
\[
\rhatadm(v)=\frac{N_{\mathrm{admitted}}(v)}{N},
\qquad
\rhattask(v)=\frac{N_{\mathrm{verified}}(v)}{N_{\mathrm{admitted}}(v)}.
\]
The search objective is
\[
g(v)=|\rhattask(v)-\rho|,
\]
only when \(\rhatadm\ge\rho_{\min}^{\mathrm{adm}}\). The v0.2 contract uses an admission floor of 0.05. Below that floor the run terminates with \code{degenerate\_gap\_admissibility\_floor}, selects no configuration, and reports \(\rhattask\) as undefined rather than fabricating zero.

This distinction is central. A value of zero means measured task failure among admitted outputs. ``Undefined'' means the task layer was never reached.

\subsection{Parameter space}

The v0.2 environment defines nine dials, eight tied directly to BoundaryBench task families and one global action-space distractor. The Cartesian grid contains exactly 2,218,750,380 points.

\begin{table}[H]
\centering
\small
\caption{BeTaL-GBI v0.2 parameter space.}
\begin{tabularx}{\linewidth}{>{\raggedright\arraybackslash}X c c c >{\raggedright\arraybackslash}p{0.29\linewidth}}
\toprule
Parameter & Min & Max & Step/count & Operational interpretation\\
\midrule
patient identity normalization & 0 & 1 & 0.05 & identity ambiguity/noise\\
orphan rate & 0 & 0.5 & 0.05 & orphan/duplicate burden\\
field anomaly bleed & 0 & 0.5 & 0.05 & structured/free-text contamination\\
code-system version validation & 0 & 1 & 0.05 & terminology/version difficulty\\
mapping arity & 1 & 6 & 6 values & RPMS-to-FHIR mapping branching\\
temporal ambiguity & 0 & 1 & 0.05 & active/historical ambiguity\\
evidence sufficiency & 0 & 10 & integer & evidence availability burden\\
policy conflict depth & 0 & 4 & integer & depth of interacting policy constraints\\
distractor actions & 0 & 5 & integer & global action-space distractors\\
\bottomrule
\end{tabularx}
\end{table}

The FHIR-facing task-family language follows the published FHIR R4 RESTful resource and transaction model \cite{fhirR4}.

The finite size is large enough that exhaustive search is unnecessary and expensive, while still giving the environment an inspectable, versioned design space. The implemented search explores a small path and local coordinate neighborhood; it does not pretend to cover all 2.2 billion points.

\begin{figure}[H]
\centering
\includegraphics[trim={1cm 2cm 1cm 2cm},  clip,  width=0.94\linewidth]{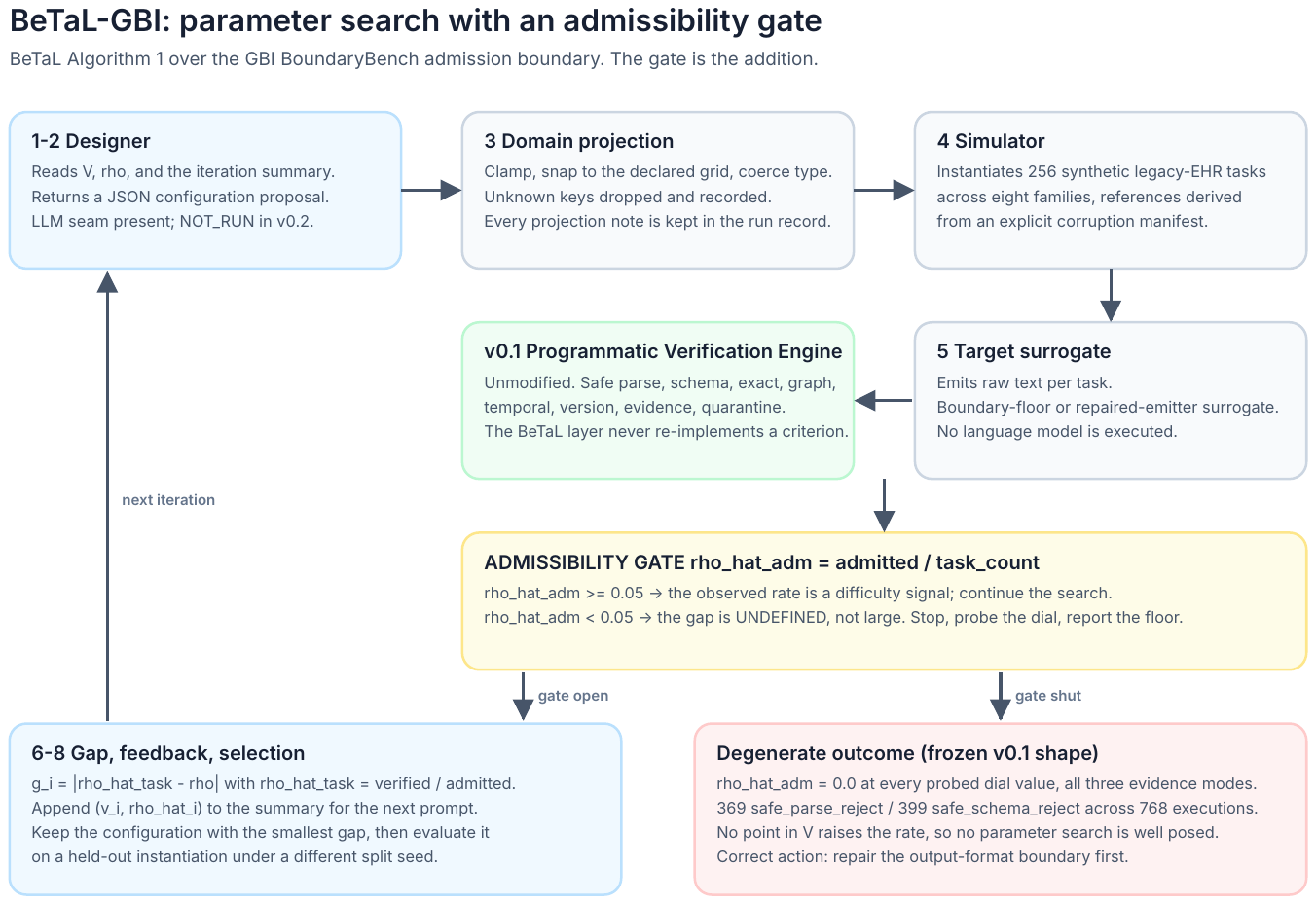}
\caption{BeTaL-GBI loop supplied with the v0.2 artifact. The important GBI-specific modification is the admission gate before task-level difficulty is interpreted.}
\label{fig:betal-loop}
\end{figure}

\subsection{Target surrogates and role separation}

The v0.2 artifact uses three transparent target types:

\begin{enumerate}
  \item \code{v01\_boundary\_floor\_surrogate}, reproducing the shape of the frozen parse/schema floor. It is not Qwen and makes no language-model claim.
  \item \code{repaired\_emitter\_cNNN}, a schema-valid declared response function with tunable competence.
  \item \code{reference\_oracle}, derived from the synthetic reference manifest and used as an oracle control.
\end{enumerate}

The designer, target, and verifier are kept separate. This prevents a designer proposal from silently becoming ground truth and prevents target competence from being confused with benchmark difficulty.

\subsection{Degenerate-gap result}

Five configurations spanning the parameter space were probed under all three v0.1 evidence modes. Every probe produced \(\rhatadm=0\). The three search attempts each halted after one of ten budgeted iterations, selected no configuration, and reported \(\rhattask\) as undefined. The per-mode synthetic boundary-floor split exactly matches the frozen v0.1 emission split: 123 parse rejects and 133 schema rejects per mode, or 369 and 399 across 768 executions.

The result is not that the benchmark is ``too hard.'' It is that difficulty is not observable until the interface failure is repaired.

\subsection{Search after format admission is restored}

Four target task-performance levels are used:
\[
\rho\in\{0.25,0.50,0.75,0.90\},
\]
labeled hard, medium, easy, and trivial. Each run uses 256 tasks and ten search iterations. The selected configuration is then re-instantiated under a different split seed for held-out measurement.

\begin{table}[H]
\centering
\caption{Mean gap across the four target levels. ``Feedback'' is the model-free feedback-coordinate implementation; the other two strategies are reimplemented baseline search strategies.}
\begin{tabular}{lrrr}
\toprule
Strategy & Mean search gap & Mean held-out gap & Held-out SD across levels\\
\midrule
Feedback coordinate & \textbf{5.23\%} & \textbf{2.87\%} & 2.30 pp\\
Random sampling + PPR & 20.90\% & 13.61\% & 8.68 pp\\
Best-of-N & 23.72\% & 11.46\% & 17.19 pp\\
\bottomrule
\end{tabular}
\label{tab:betal-mean}
\end{table}

\begin{table}[H]
\centering
\caption{Held-out target gaps by difficulty level. Best-of-N wins two individual cells; feedback has the best mean and is substantially better at the high-performance edge.}
\begin{tabular}{lrrrr}
\toprule
Level & Target \(\rho\) & Feedback & Random+PPR & Best-of-N\\
\midrule
Hard & 0.25 & 0.78\% & 6.25\% & \textbf{0.39\%}\\
Medium & 0.50 & 1.17\% & 6.64\% & \textbf{0.00\%}\\
Easy & 0.75 & \textbf{3.91\%} & 17.58\% & 8.98\%\\
Trivial & 0.90 & \textbf{5.63\%} & 23.98\% & 36.48\%\\
\bottomrule
\end{tabular}
\label{tab:betal-levels}
\end{table}

\begin{figure}[H]
\centering
\includegraphics[trim={1cm 2cm 1cm 2cm},  clip,  width=0.90\linewidth]{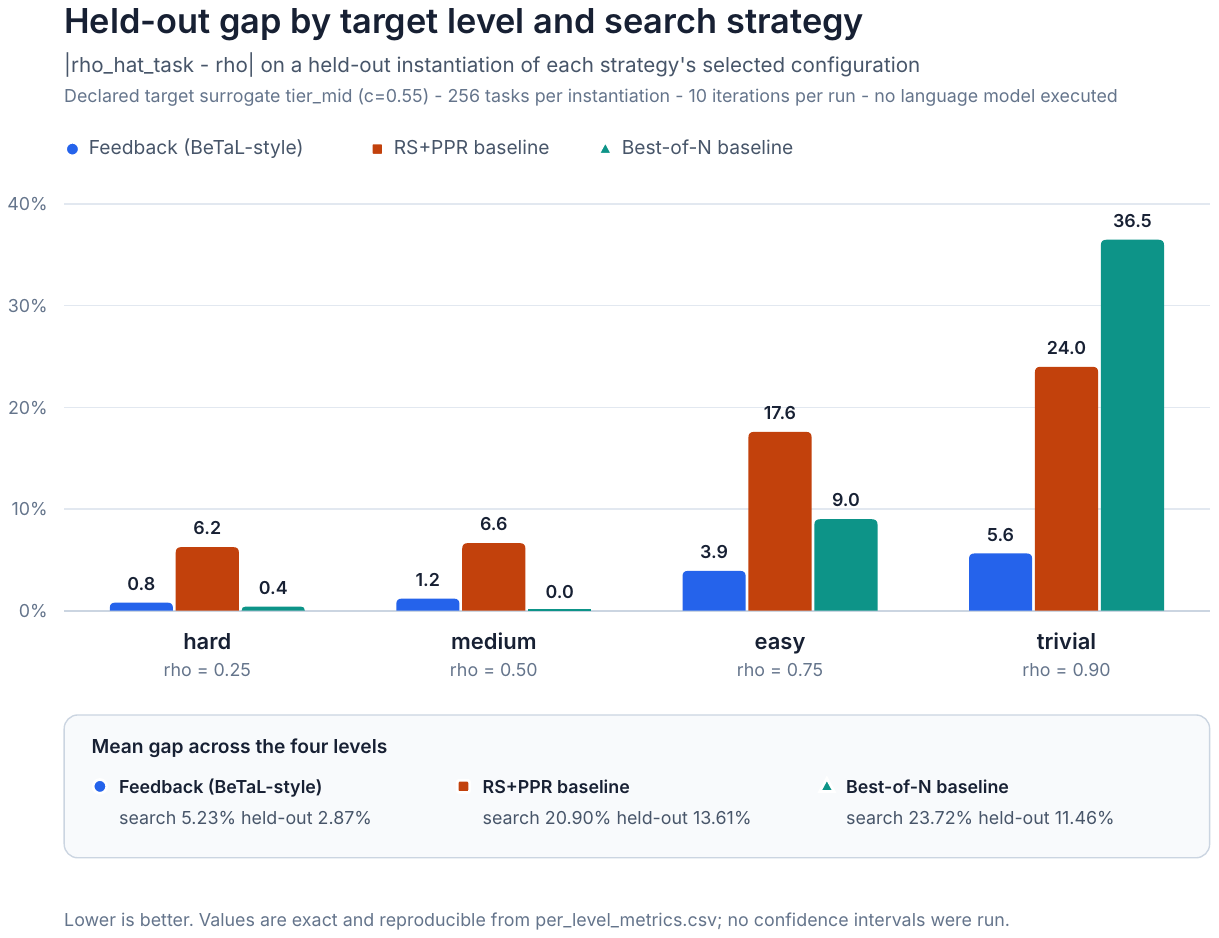}
\caption{Held-out gap comparison from the supplied v0.2 artifacts. These are simulator/harness results; no LLM designer or target was executed.}
\label{fig:betal-gap}
\end{figure}

The result is qualitatively consistent with the motivation for feedback-driven benchmark design: random search can land near central targets by chance, while feedback becomes more useful near compressed edges of the reachable range. The artifact also exposes a limitation of the deterministic reference designer: in at least one hard-level run, several late iterations repeat the same proposal and therefore waste search budget.

\subsection{Reachability and monotonicity are properties of the space, not the optimizer}

A target can only be achieved if the parameterized environment can express it for the selected target system. The monotone-dial probe reports the following empirical ranges:

\begin{table}[H]
\centering
\small
\caption{Observed task-performance ranges on the declared monotone dial. A positive floor is a property of the reachable space, not necessarily an optimizer failure.}
\begin{tabular}{lccccc}
\toprule
Target tier & Range & 0.25 & 0.50 & 0.75 & 0.90\\
\midrule
Low (c=0.35) & 0.016--0.727 & reachable & reachable & floor 2.34\% & floor 17.34\%\\
Mid (c=0.55) & 0.121--0.898 & reachable & reachable & reachable & floor 0.16\%\\
High (c=0.75) & 0.238--0.941 & reachable & reachable & reachable & reachable\\
\bottomrule
\end{tabular}
\end{table}

Across 24 consecutive dial steps, two local monotonicity violations occur. At \(N=256\), the artifact attributes them to finite-sample variation rather than silently forcing a monotone fit. The important governance behavior is that monotonicity is tested rather than assumed.

\subsection{A genuine mapping cone, still diagnostic-only}

The original architecture manuscript carefully labels its appendix calculation as a projector-based obstruction surrogate. BeTaL-GBI v0.2 adds a finite sheaf morphism \(\phi:F\to G\), constructs the corresponding mapping-cone differential using standard cellular-sheaf and homological-algebra constructions \cite{hansen2019sheaves,weibel1994homological}, and checks the chain condition
\[
d^0d^{-1}=0.
\]
A rank-deficient restriction \(\operatorname{diag}(1,0)\) is required; identity restrictions collapse the intended degrees of freedom. Across all eight binary agreement patterns, the implementation obtains
\[
\dim H^0(\operatorname{Cone}\phi)=\#\{\text{disagreeing axes}\},
\]
with trace energy equal to one on disagreeing axes and zero on agreeing axes.

This is a stronger numerical implementation than the original projector surrogate, but its scoring weight remains zero and the diagnostic is explicitly not clinically validated. The evaluation therefore demonstrates algebraic behavior, not clinical correctness.

\section{GBI v2: from refusal to selective external-validity gating}

\subsection{Why v0.1 was insufficient as a positive substrate result}

At zero coverage, ``severe contradiction sensitivity = 100\%'' is vacuous because every item is refused. Likewise, no selective-risk estimate exists. GBI v2 is designed to test whether the gate can simultaneously do two things:

\begin{enumerate}
  \item refuse declared severe contradictions, including adversarial proposals that are syntactically valid; and
  \item allow clean, policy-compatible work to proceed rather than treating refusal itself as success.
\end{enumerate}

The v2 evaluation uses 512 synthetic tasks over the same eight task families. Its runtime gate is grounded in \(W\) and \(P\), while the hidden reference remains an after-the-fact measurement instrument.

\subsection{Deterministic format repair: admission without invented correctness}

A deterministic repair stage performs only declared syntactic transformations: code-fence stripping, balanced-object extraction, bare-key quoting, duplicate-key collapse with first value retained, schema-version insertion, a small action-synonym table, and extra-key dropping. It deliberately refuses to fabricate content for truncated payloads or non-object answers.

\begin{table}[H]
\centering
\caption{Effect of deterministic repair on the frozen v0.1-style format-floor target.}
\begin{tabular}{lrr}
\toprule
Quantity & Frozen floor & After repair\\
\midrule
Admissibility rate & 0.0000 & \textbf{0.4648}\\
Coverage & 0.0000 & \textbf{0.4473}\\
Selective risk & undefined & \textbf{1.0 (defined)}\\
Verified completion & 0.0000 & \textbf{0.0000}\\
\bottomrule
\end{tabular}
\end{table}

This is an important negative control. Repair makes malformed output measurable; it does not make wrong output correct. Sixty truncated payloads and seventy non-object answers remain unrepaired because recovering them would require inventing semantic content.

\subsection{Severe-contradiction injection}

The v2 population contains 116 injected severe contradictions across ten classes, disjoint from 99 records intentionally left clean for the narrow false-conflict denominator. Four classes are absent from the base simulator and therefore test additional gates.

\begin{table}[H]
\centering
\small
\caption{Injected severe contradiction classes.}
\begin{tabular}{lr@{\hspace{2em}}lr}
\toprule
Class & Count & Class & Count\\
\midrule
Identity ambiguous & 12 & Identity unresolvable & 12\\
Provenance absent & 11 & Required evidence absent & 11\\
Structured field fully contaminated & 11 & Terminology unpinned & 12\\
Terminology unsigned & 12 & Terminology unresolvable & 11\\
Validity expired & 12 & Validity invalid & 12\\
\midrule
\multicolumn{3}{r}{Total} & 116\\
\bottomrule
\end{tabular}
\end{table}

\subsection{Policy conformance and non-vacuity}

The declared v2 policy contains 16 gates. Direct probes exercise each gate, and a clean witness is admitted as a non-vacuity control. Two intentionally defective policies are also evaluated:

\begin{itemize}
  \item an incomplete ablation omits two gates and incorrectly admits an unsigned-terminology probe;
  \item a score-only identity ablation omits ambiguity and unresolvable-identity gates and fails the corresponding probes.
\end{itemize}

This is a load-bearing test: if the ablated policies produced the same result as the unablated declared policy, the positive score would say little about whether the omitted gates are load-bearing.

\subsection{Adversaries that clear the format boundary}

Two declared synthetic adversaries directly test the manuscript's containment claim.

\begin{table}[H]
\centering
\small
\caption{GBI v2 adversaries. These are deterministic surrogates, not language models.}
\begin{tabularx}{\linewidth}{>{\bfseries}p{0.23\linewidth} X c c}
\toprule
Adversary & Attack & Admitted & Silent promotions\\
\midrule
Confident hallucinator & Emits format-valid content, confidence 0.97, correct evidence references, and always requests admission. It makes 221 attempted over-admissions. & 98 & \textbf{0}\\
Evidence forger & Proposes the witness-consistent action but cites one nonexistent evidence reference and an unwitnessed source record. & 0/512 & \textbf{0}\\
\bottomrule
\end{tabularx}
\end{table}

The pair attacks independent dimensions. A gate that checked only action plausibility could miss the forger; a gate that checked only citations could miss an over-admitting proposal. Zero silent promotions is measured against both witness severity and the synthetic reference action. The result is evidence for the implemented gate under these declared attack classes, though not proof against arbitrary hallucinations or adaptive attackers.

\subsection{Clinical-style scorecard and denominator discipline}

Under a predeclared selection rule -- the largest strictness value satisfying both measurable Table 3 clinical-style targets -- the declared policy selects strictness 0.6.

\begin{table}[H]
\centering
\small
\caption{GBI v2 scorecard against the architecture manuscript's proposed Table 3 targets.}
\scriptsize
\begin{tabularx}{\linewidth}{>{\RaggedRight\arraybackslash}p{0.15\linewidth} >{\RaggedRight\arraybackslash}X >{\RaggedRight\arraybackslash}p{0.20\linewidth} >{\RaggedRight\arraybackslash}p{0.16\linewidth}}
\toprule
Group & Measure & Result & Evidence class\\
\midrule
Mathematical & Spectral gap \(\lambda_1-\lambda_0\ge0.15\) on genuine cone Laplacians & 1.0000 (within floating point) & MEASURED\\ [0.6em]  \vspace{-0.4em} \\
Mathematical & Fisher condition number \(\le 10^4\) over corrected evidence box & 9998.3472 & MEASURED\\ [0.6em]  \vspace{-0.4em} \\
Systems & End-to-end enclave latency \(\le150\) ms & 0.0968 ms CPU-only software path & PARTIAL PROXY\\ [0.6em]  \vspace{-0.4em} \\
Systems & Attestation bootstrapping \(\le2.5\) s & not measured & OUT OF SCOPE\\ [0.6em]  \vspace{-0.4em} \\
Clinical-style & Severe contradiction sensitivity = 100\% & 116/116 & MEASURED SYNTH.\\ [0.6em] \vspace{-0.4em}  \\
Clinical-style & False conflict adjudication \(\le4\%\) & 0/99 = 0.0\% narrow & MEASURED SYNTH.\\
\bottomrule
\end{tabularx}
\end{table}

Five of five measurable targets are met; one requires a real TEE/attestation service and is left out of scope. The latency number is not an enclave measurement. It is a lower-scope CPU-only timing of deterministic checks and therefore cannot be used as evidence that a hardware enclave deployment meets 150 ms.

\begin{figure}[H]
\centering
\includegraphics[trim={3cm 4cm 3cm 4cm},  clip,  width=0.86\linewidth]{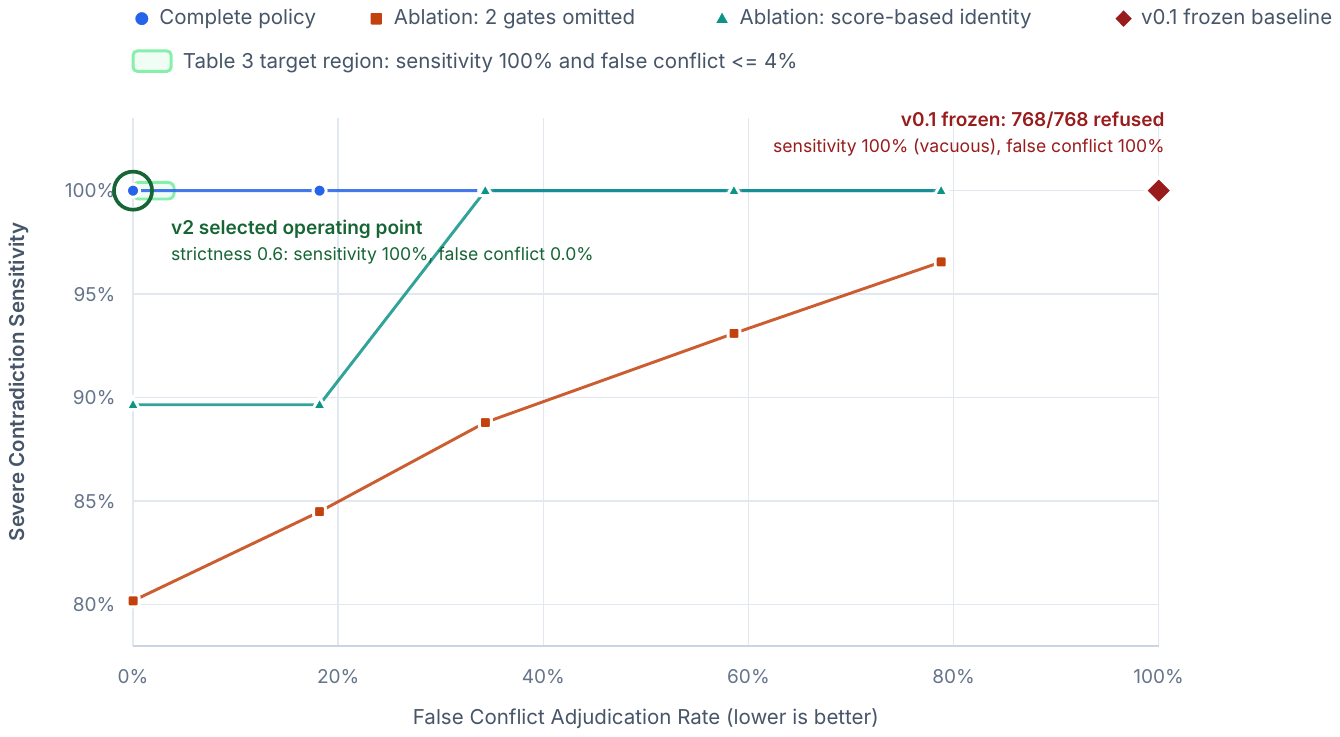}
\caption{Sensitivity/false-conflict frontier from the v2 artifact. The frozen v0.1 point has 100\% sensitivity only because every record is refused; its false-conflict rate is therefore 100\%.}
\label{fig:frontier}
\end{figure}

The false-conflict result has a denominator sensitivity that must not be hidden. The narrow, predeclared denominator contains the 99 records deliberately left clean by injection and produces 0.0\%. A pessimistic broader denominator yields 5/117 = 4.27\%, slightly missing the 4\% proposed target. Both figures remain part of the result. Figure~\ref{fig:frontier} shows the strictness frontier and policy ablations.

\subsection{Strictness and the visible operating boundary}

The declared policy satisfies the two narrow clinical-style targets from strictness 0.0 through 0.6. At strictness 0.7, the clean-record false-conflict rate rises to 18.18\%, so the selection rule has an observable boundary rather than choosing an arbitrary round number. The policy ablations fail in diagnostically different ways.

This is the desired behavior for a benchmarked enterprise policy: strictness should expose a tradeoff, and an operating point should be derived from a declared rule rather than selected after viewing the result.

\subsection{Surgical quarantine is a dependency-scope statement}

The v2 liveness experiment asks how much independently admissible work survives when one record is refused.

\begin{table}[H]
\centering
\caption{Liveness under four quarantine-closure semantics.}
\begin{tabular}{lrr}
\toprule
Quarantine scope & Admissible work surviving & Liveness\\
\midrule
Record-scoped refusals only & 112/112 & \textbf{1.000}\\
Declared scope incl. mandated shared freezes & 84/112 & 0.750\\
Naive closure over every shared reference & 84/112 & 0.750\\
Coarse family-level closure & 0/112 & \textbf{0.000}\\
\bottomrule
\end{tabular}
\end{table}

Twenty-four mandated administrative freezes close over nine shared scopes because Boundary 2 requires an unsigned or unpinned shared terminology bundle to halt dependent use. By contrast, a bad record that merely cites a valid shared bundle does not taint the bundle itself. ``Surgical'' therefore means dependency-aware pre-commit isolation, never partial success inside an operation whose underlying transaction is atomic.

\section{Mathematical validation and the discovered Fisher-bound erratum}

\subsection{Appendix-style validation}

GBI v2 expands the architecture manuscript's proposed Appendix B.1 checks:

\begin{enumerate}
  \item A finite Boolean algebra with 64 elements over six atoms is stress-tested over 65,536 randomized operations, checking eleven laws per operation with zero violations.
  \item The Dirichlet Fisher metric is evaluated at all 16 corners of the declared four-dimensional evidence box plus 512 adversarial near-boundary probes.
  \item The genuine mapping-cone implementation evaluates all eight agreement patterns, verifies the chain condition, symmetry and positive semidefiniteness to numerical tolerance, and checks basis-rotation invariance of trace energies.
\end{enumerate}

For the cone suite, the worst symmetry residual is 0, the numerical minimum eigenvalue is approximately \(-5.39\times10^{-32}\), and the worst energy rotation drift is approximately \(1.11\times10^{-15}\), all consistent with floating-point error.

\subsection{The one-dimensional bound was not a box-wide guarantee}
\label{sec:fisher-erratum}

The architecture manuscript observed that, along the slice
\[
\alpha=(\epsilon,3,4,5),
\]
the condition-number budget \(\kappa_2(g)\le10^4\) is crossed near \(\epsilon\approx0.066\). The v3 independent recomputation solves this slice as
\[
\epsilon_{\mathrm{slice}}=0.066021703.
\]
That number is correct for the slice.

However, Appendix B.1 asks for a corner sweep of a declared evidence box. For \([\epsilon,20]^4\), the worst corner is not the slice above. At \(\epsilon=0.066021703\), the worst-corner condition number is approximately
\[
4.79\times10^5,
\]
about 48 times the declared budget. The box-wide lower bound that satisfies the budget at ceiling \(A=20\) is
\[
\boxed{\epsilon_{\mathrm{box}}\approx0.326472}.
\]
The required \(\epsilon\) depends strongly on the ceiling:

\begin{table}[H]
\centering
\caption{Evidence-box lower bound required for the \(\kappa_2\le10^4\) budget.}
\begin{tabular}{rrr}
\toprule
Ceiling \(A\) & Box-wide \(\epsilon\) & One-dimensional slice value\\
\midrule
5 & 0.093001 & 0.066022\\
10 & 0.161794 & 0.066022\\
20 & \textbf{0.326472} & 0.066022\\
50 & 0.918946 & 0.066022\\
\bottomrule
\end{tabular}
\end{table}

This is an erratum to a stated numerical implication, not a refutation of the continuity argument that a compact evidence box has finite conditioning bounds. The corrected v2/v3 scorecard uses the box-wide value.

\begin{claimbox}
\textbf{Why the erratum matters scientifically.} A verification framework earns credibility not only by passing tests but by making its own failed assertions expensive to hide. The v3 claim register classifies this item as \texttt{ERRATUM}, preserves the original statement, records the corrected bound, and excludes it from the ``met'' count.
\end{claimbox}

\section{GBI-DCSE v3: evaluating the full testable architecture}

\subsection{Claim-register methodology}

The v3 evaluation enumerates 99 claims from the architecture manuscript and binds each to section, evidence class, resolver, expected behavior, and artifact evidence. The register covers manuscript Sections 1--12 and Appendices A--B.

The aggregate result is:

\begin{table}[H]
\centering
\caption{GBI-DCSE v3 claim coverage.}
\begin{tabular}{lr}
\toprule
Quantity & Count\\
\midrule
Total registered claims & 99\\
Testable claims & 96\\
Testable claims met & \textbf{95}\\
Testable claims unmet & 1 (erratum)\\
Explicitly out of scope & 3\\
Declared verification checks & \textbf{148}\\
Verification failures & \textbf{0}\\
\bottomrule
\end{tabular}
\end{table}

The evidence-class distribution is 16 REPRODUCED, 51 MEASURED, 5 MEASURED SYNTH., 22 STRUCTURAL, 1 PARTIAL PROXY, 3 OUT OF SCOPE, and 1 ERRATUM.

\begin{figure}[H]
\centering
\includegraphics[trim={1cm 2cm 1cm 2cm},  clip,  width=0.92\linewidth]{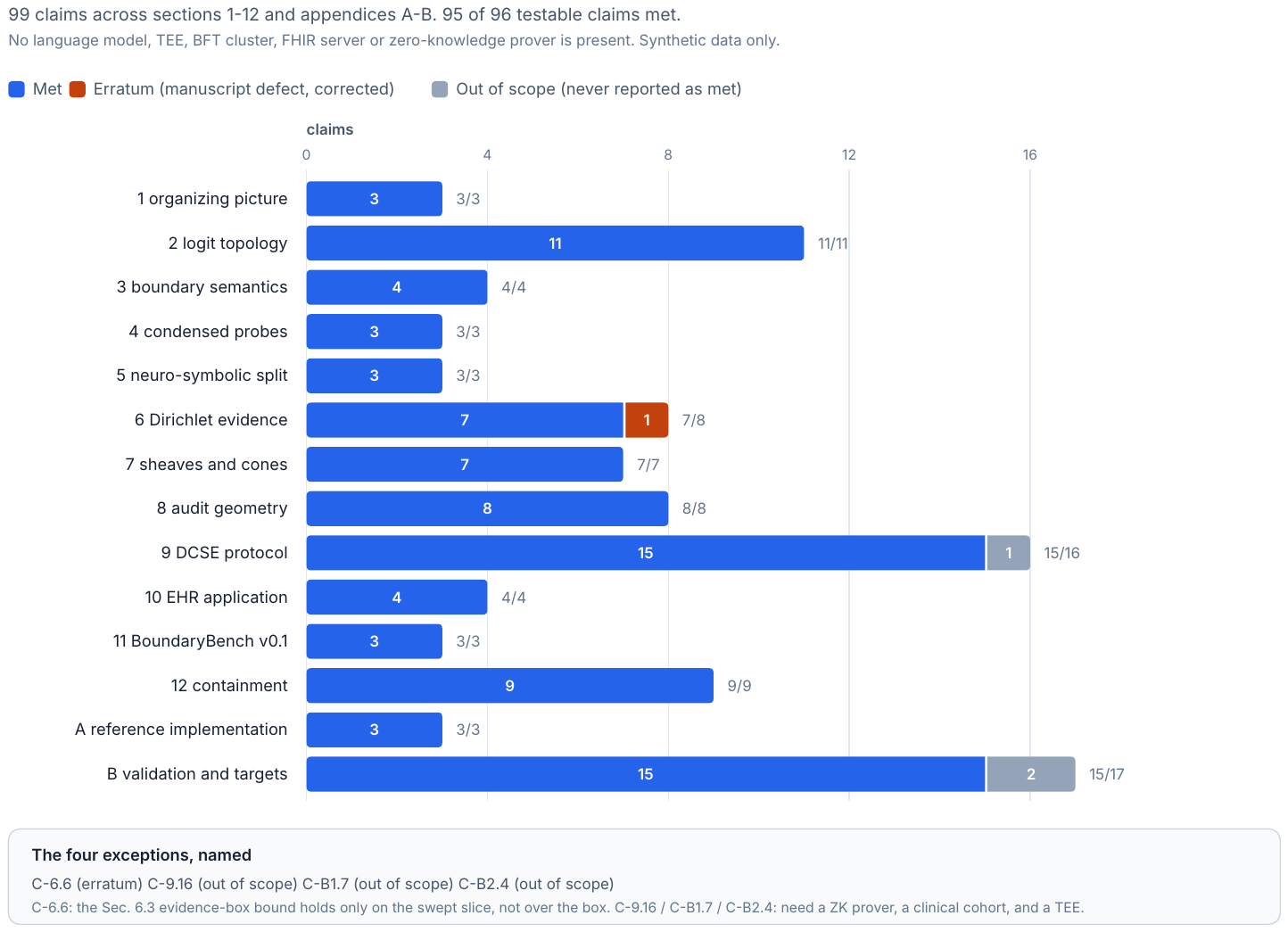}
\caption{Claim coverage generated from the v3 claim register. The presence of explicit OUT OF SCOPE and ERRATUM classes is part of the evaluation design, not an afterthought.}
\label{fig:v3-coverage}
\end{figure}

The register's own integrity is tested. A non-out-of-scope claim may not silently resolve to a missing value, out-of-scope claims may not be reported as met, identifiers must be unique, and counts must reconcile. The only unmet testable claim is the Fisher-bound erratum described above.

The two headline counts answer different questions. ``95/96 testable claims met'' scores the registered scientific/system assertions against their declared criteria. ``148/148 verifier checks passed'' scores whether the verification harness correctly reproduced values, exercised faults, enforced scope classifications, and detected expected failures. A check that correctly identifies the Fisher statement as an \texttt{ERRATUM} is therefore a passing verifier check, not evidence that the original Fisher statement was correct.

\subsection{Appendix A and previously untested GBI claims}

The v3 evaluator ports and independently recomputes the numerical appendix rather than merely comparing serialized values. Eleven self-check assertions pass and 21 published numerical values reproduce. Examples include
\[
H=1.701632,\qquad J=1.028000,\qquad K_O=1.922661,
\]
and the moderate-interior Fisher condition number \(\kappa_2\approx20.457\). The trigamma approximation agrees with an independent series/asymptotic implementation to a maximum error of approximately \(2.49\times10^{-9}\).
\newpage

Additional checks cover claims that the original executable appendix did not implement:

\begin{itemize}
  \item exact affine-reconstruction residual below \(2.2\times10^{-15}\);
  \item probe-visible kernel perturbations up to 1000 times scale with output drift below \(10^{-12}\), while a visible control direction produces substantial drift;
  \item all four softmax-entropy rows reproduce to six decimals and top-3 tail mass is 0.041708;
  \item hard truncation produces infinite \(D_{\mathrm{KL}}\) when positive full-support mass is assigned to removed categories;
  \item exhaustive Boolean-homomorphism checks over 4,096 pairs produce zero violations;
  \item a driven two-state system crosses category boundaries twice while entropy remains at least 0.665 nats, 96\% of the two-category maximum \(\ln 2\);
  \item all four higher-dimensional chart safety checks are exercised; and
  \item Boundary 3 is tested in both directions so visualization quality cannot become admission authority.
\end{itemize}

These tests close gaps in the first paper's executable coverage without changing the paper's nonclaims: geometric charts remain advisory and no numerical chart metric independently establishes real-world truth.

\section{The DCSE systems layer}

\subsection{Identity/provenance ledger with cryptographic signatures}

The v3 ledger uses cryptographic Ed25519 signatures, as specified in RFC 8032 \cite{rfc8032}, via the Python \code{cryptography} package, together with SHA-256 digests. Seven fault classes are injected and correctly detected/classified. A clean ledger remains non-equivocating, while every declared fallback trigger halts authoritative writes. A clean fast path permits writes, providing the necessary non-vacuity control.

Equivocation evidence is publicly checkable from the signed conflicting records and public key. A tampered signature fails verification. Crucially, the test suite also constructs a ledger that is structurally valid but binds the wrong identity, preserving the distinction:

\[
\text{ledger consistency} \not\Rightarrow \text{identity truth}.
\]

The ledger proves provenance/non-equivocation properties of the recorded decision process, not correctness of the underlying entity match.

\subsection{Attestation verification and fail-closed behavior}

The software attestation model accepts a valid baseline quote and tests seven failing cases:

\begin{enumerate}
  \item invalid signature;
  \item expired quote;
  \item modified measurement;
  \item correctly signed but unlisted measurement;
  \item replayed nonce;
  \item revoked platform key; and
  \item downgraded TCB version.
\end{enumerate}

Each case denies the governed write path through the intended check. The result is useful because the cases are distinguishable, rather than collapsing every failure into one catch-all branch.

There is no hardware root of trust. The measured software verification path is a proxy only; it excludes quote generation and remote attestation service round trips. Therefore the manuscript's proposed 2.5-second attestation bootstrapping target remains OUT OF SCOPE.

\subsection{Consensus model and fault boundary}

The consensus module models the classical PBFT-style fault/quorum condition \cite{castro1999pbft}
\[
n\ge3f+1,\qquad q=2f+1.
\]
For \(f=1,\ldots,5\), the verifier independently re-derives quorum arithmetic. Across 62 analyzed configurations, safety holds within the declared Byzantine tolerance, progress holds for modeled crash/partition cases up to the declared bound, and the write path halts when the active configuration falls below threshold. At \(f+1\) Byzantine participants, an unsafe split is constructible, demonstrating that the modeled bound is tight rather than merely sufficient in the tested abstraction.

This is exhaustive over the state space of the model; it is not a running BFT network, and no latency/throughput claim is made.

\subsection{Sparse enclave certificate: why spectral moments are load-bearing}

The architecture proposes keeping small deterministic checks inside a constrained enclave while moving dense linear algebra outside. The v3 implementation evaluates that split on a synthetic \(240\times240\) operator with 477 nonzeros and true obstruction dimension three.

The sparse verifier checks orthonormality/residual conditions and spectral moments derived from
\[
\sum_i\lambda_i=\operatorname{tr}(L),\qquad
\sum_i\lambda_i^2=\|L\|_F^2.
\]
An honest dense solver certificate is accepted. Nine forgery classes are rejected.

A particularly important negative control under-claims the kernel dimension, attempting to hide an obstruction. Residual checking alone accepts this forged subspace with a worst residual around \(2.28\times10^{-15}\). The spectral-moment conditions reject it. Thus the second half of the certificate is empirically load-bearing; it is not ornamental redundancy.

The declared resource accounting is:

\begin{table}[H]
\centering
\caption{Sparse versus dense verification accounting in the v3 synthetic enclave experiment.}
\begin{tabular}{lrrr}
\toprule
Resource & Sparse path & Dense path & Declared budget\\
\midrule
Floating-point operations & 8,440 & 124,416,000 & 5,000,000\\
Heap bytes & 13,440 & 921,600 & 262,144\\
Syscalls & 2 & not the enclave path & 64\\
\bottomrule
\end{tabular}
\end{table}

The dense/sparse flop ratio is approximately \textbf{14,741x}, and the heap ratio is approximately 68.6x. These are algorithmic/accounting results in ordinary software, not measured SGX/TEE performance.

\subsection{Atomic rollback and liveness can coexist}

The transaction module tests a clean initial store, successful independent transactions, and a forced transaction containing an invalid medication request. For the bad transaction:

\begin{itemize}
  \item no entry from the bad bundle is committed;
  \item the store digest is byte-identical before and after rejection;
  \item unrelated records remain unchanged;
  \item an error \code{OperationOutcome} with \code{business-rule} is emitted; and
  \item the failed attempt is recorded in the audit path.
\end{itemize}

Independent work progresses separately while held-back work is rerouted. The experiment therefore demonstrates the intended distinction: stalk/work-item quarantine occurs before transaction construction or across independently scoped transactions; an atomic transaction itself remains all-or-nothing. The use of transaction semantics and structured OperationOutcome responses is aligned with the FHIR R4 RESTful interaction model \cite{fhirR4}.

\subsection{Receipts, replay, and review surfaces}

The v3 receipt chain verifies coherence, completeness, hash linkage, and replay. Fourteen of fourteen decisions are reproduced from receipts. A tampered receipt is detected. Every receipt pins policy and terminology versions, and forbidden review-surface fields are absent.

A deliberate human-factors design choice is tested structurally: model confidence is not carried as an authoritative review field. The intended review artifact exposes the candidate, authoritative evidence, policy rule, version, provenance, failed criterion, and required action. This does not eliminate automation bias, but it avoids making a single model score the primary authorization surface.

\subsection{Consistency certificate and the zero-knowledge boundary}

The finite cone-certificate implementation accepts honest certificates for both vanishing and non-vanishing \(H^1\) and rejects false cohomology claims. The public input digest binds the policy version, so a certificate cannot be replayed unchanged across policies.

Zero knowledge is deliberately not implemented or claimed. A binding, blinded commitment and finite verification predicate establish prerequisites for future ZK work, but they are not a zero-knowledge proof system. This is one of the three v3 OUT OF SCOPE claims.
\newpage

\section{Cross-domain portability: measured code reuse, not semantic equivalence}

\subsection{What changes and what remains invariant}

The architecture manuscript argues that GBI/DCSE can be specialized to healthcare, financial services, and government/mission systems by changing domain semantics while retaining the verification contract \cite{spivey2026gbi}. V3 tests this proposition by instantiating a synthetic sensitive-infrastructure domain.
Eight modules are reused verbatim:
\begin{center}
\code{crypto}, \code{ledger}, \code{attestation}, \code{consensus}, \code{enclave}, \code{transaction}, \code{receipts}, \code{cone\_certificate}.
\end{center}
Four domain-specific objects are replaced:
\begin{center}
\code{InfrastructureWitness}, \code{required\_action}, \code{GATE\_PRECEDENCE}, \code{SEVERE\_CLASSES}.
\end{center}
No architectural change is required. This is evidence of implementation portability of the declared systems layer.

\subsection{Synthetic infrastructure population}

The population contains 512 synthetic records. Of these, 265 carry one of ten severe contradiction classes, 175 are fully clean, and the remainder require legitimate review or historical-only release. Domain-specific gates add classification-handling caveats, need-to-know, jurisdiction, source-reliability rating, and directive-version pinning to shared identity, signature, provenance, and validity checks.

\begin{table}[H]
\centering
\caption{Sensitive-infrastructure synthetic triage result.}
\begin{tabular}{lr}
\toprule
Measure & Result\\
\midrule
Severe contradiction sensitivity & \textbf{265/265 = 1.000}\\
Severe contradictions missed & \textbf{0}\\
False conflicts over fully clean records & \textbf{0/175 = 0.000}\\
Receipt chain & valid\\
Ledger & valid and non-equivocating\\
Forced-bad-entry atomicity & held\\
Unrelated records & unchanged\\
\bottomrule
\end{tabular}
\end{table}

The resulting action distribution is 175 release-verified, 54 release-historical-only, 157 reject, 78 quarantine-record, 44 analyst-review, and 4 abstain.

The defensible interpretation is that the same systems modules can act as a router/triage substrate when the witness schema and policy are rewritten for a new domain. The result does not establish readiness for classified, financial, clinical, or other real infrastructure. No real assets, operators, locations, telemetry, authoritative registries, or reviewed handling-caveat policies are present.
\section{What changed from v0.1 to v3}

The four stages answer different questions and should not be merged into one headline score.

\small
\begin{longtable}{>{\RaggedRight\arraybackslash}p{0.10\linewidth} >{\RaggedRight\arraybackslash}p{0.24\linewidth} >{\RaggedRight\arraybackslash}p{0.28\linewidth} >{\RaggedRight\arraybackslash}p{0.28\linewidth}}
\caption{Evaluation progression and what each stage licenses.}\\
\toprule
Stage & Main question & Positive evidence & What it does not establish\\
\midrule
\endfirsthead
\toprule
Stage & Main question & Positive evidence & What it does not establish\\
\midrule
\endhead
v0.1 & Will a frozen model output cross a strict contract? & 768 executions all fail closed at parse/schema. & Selective utility, task difficulty, policy quality, general model safety.\\
\addlinespace
v0.2 & Can benchmark difficulty be tuned without confusing interface failure with task failure? & Degenerate objective detected at zero admission; after repair, feedback search reaches 2.87\% mean held-out target gap. & LLM-designer performance, commercial-model difficulty, full-space optimality.\\
\addlinespace
v2 & Can the GBI gate selectively admit clean work and refuse declared severe contradictions without answer-key leakage? & 116/116 severe injections caught; 0/99 narrow clean false conflicts; zero silent promotions under two synthetic adversaries; policy ablations fail. & Real clinical safety, complete taxonomy, adversarial robustness to arbitrary attacks.\\
\addlinespace
v3 & Do the broader testable manuscript claims have executable evidence? & 95/96 testable claim criteria met; 148 declared checks; systems modules, receipts, rollback and portability exercised. & Real TEE/BFT/FHIR/ZK deployment, real data, operational accreditation.\\
\bottomrule
\end{longtable}

A key methodological shift is visible across the sequence. V0.1 measures the \emph{model-to-contract boundary}. V0.2 measures the \emph{benchmark-design environment}. V2 measures the \emph{reference-independent policy gate}. V3 measures the \emph{claim-to-evidence relationship of the full architecture}. A mature evaluation program needs all four views.

\section{Threat model and residual exclusions}
\label{sec:threat-model}

The definitive evidentiary perimeter is Table~\ref{tab:system-boundaries}; it is not re-litigated in every technical section. Within that perimeter, the implemented threat model is \emph{bounded adversarial and integrity testing of a declared verification substrate}. The evaluated attacks include malformed emissions, format-valid over-admission attempts, forged evidence references, policy ablations, ledger faults, attestation faults, quorum failures, certificate forgeries, replay/tamper attempts, and invalid entries in an otherwise atomic transaction.

Three residual categories are especially important for further review. First, the adversaries are declared and non-adaptive; they do not search the implementation for unknown policy gaps or exploit side channels. Second, the systems tests assume the integrity of roots that are only modeled here, including institutional authority, key provisioning, and hardware trust. Third, the verifier can only enforce predicates represented in \(W\), \(V\), and \(P\); incomplete source systems or omitted policy semantics remain external failure modes.

These exclusions do not weaken the reported synthetic results; they define the conditions under which those results are valid and the experiments required to extend them. Section~\ref{sec:introduction} therefore treats scope as part of the evaluation contract rather than as a sequence of defensive footnotes.
\newpage

\section{Relation to benchmark and runtime-verification literature}

GBI-DCSE sits at the intersection of several literatures rather than replacing any one of them.

\subsection{Dynamic benchmark design}

BeTaL parameterizes benchmark environments and uses an LLM designer to target properties such as difficulty \cite{dsouza2025betal}. The GBI adaptation adds a precondition that is specific to structured enterprise evaluation: task performance is not optimized when the output interface is not admitted. BenchBench independently highlights the need to evaluate benchmark-generation quality and invalidity rather than treating generated benchmark items as automatically valid \cite{zheng2026benchbench}. Together these works motivate benchmark design as an object of evaluation in its own right.

\subsection{Agent benchmarks and policy-following}

\(\tau\)-bench evaluates tool-using agents in dynamic user interactions with domain-specific policies and scores the resulting database state \cite{yao2024taubench}. GBI-DCSE is complementary: it focuses on the pre-action admissibility boundary, versioned evidence, and explicit quarantine/review semantics rather than only end-state task success.

\subsection{Formal monitoring and intervention}

Recent work combines formal temporal logic with offline auditing, online monitoring, prediction, and runtime intervention for black-box AI systems \cite{alamdari2026formal}. That line provides stronger formal treatment of temporally extended behavioral constraints than the current GBI implementation. GBI/DCSE contributes a different systems decomposition centered on enterprise evidence, policy versions, provenance, identity, dependencies, and transaction routing.

\subsection{Healthcare interoperability evaluation}

FHIR-AgentBench evaluates LLM agents on realistic FHIR-grounded clinical question answering and retrieval strategies \cite{lee2025fhiragentbench}. BoundaryBench targets a different question: whether proposed transformations or actions are admissible under explicit enterprise constraints. FHIR-specific semantics therefore motivate one instantiation of the boundary but are not themselves the claimed novelty of the architecture.

\section{Implications for frontier-evaluation and enterprise-AI programs}

The results suggest several research directions that matter beyond healthcare.

\subsection{Failure surfaces should be factored by layer}

A single ``accuracy'' number conflates at least four failure surfaces:
\begin{enumerate}
  \item interface/format admission;
  \item task-semantic correctness among admitted outputs;
  \item evidence/policy admissibility for downstream action; and
  \item systems integrity of the control plane executing that decision.
\end{enumerate}

The v0.1-to-v3 sequence shows why this factorization matters. Optimizing task difficulty at zero admission is meaningless. Measuring sensitivity while refusing everything is vacuous. Measuring a policy without ablations can hide non-load-bearing gates. Measuring a sparse certificate without an adversarial under-claimed kernel can leave the critical half of the certificate untested.

\subsection{A benchmark can become an enterprise diagnostic loop}

A domain-specific deployment can be conceptualized as
\[
\text{customer stack}
\rightarrow \text{local evaluation}
\rightarrow \text{failure surface}
\rightarrow \text{failure slices}
\rightarrow \text{intervention}
\rightarrow \text{re-evaluation}.
\]
The intervention might be expert labeling, synthetic/counterfactual data, evaluator development, retrieval changes, policy refinement, structured-output engineering, post-training, or systems repair. The benchmark should not predetermine which intervention or vendor is correct; its value is the reproducible diagnosis.

\subsection{Claim registers are useful for research-to-production translation}

A 99-row claim register is not merely documentation. It forces a research program to state which claims are reproducible numbers, synthetic measurements, structural properties, proxies, missing experiments, or errors. This provides a practical bridge between paper prose and an engineering acceptance plan.

For enterprise deployment, the same pattern could be extended to customer-specific controls:
\[
\text{policy rule}\leftrightarrow\text{test}\leftrightarrow\text{evidence}\leftrightarrow\text{version}\leftrightarrow\text{owner}.
\]
That form is especially useful when a model, evaluator, policy, and source system can change on different schedules.

\section{Discussion}

\subsection{Self-falsification is an acceptance test, not a blemish}

The Fisher correction provides the clearest reason to treat the claim register as part of the scientific contribution. A conventional validation appendix can become confirmatory if its tests are written only to reproduce numbers already believed to be correct. Here, the stronger box-wide test contradicts a published numerical implication, and the reporting machinery is required to preserve that contradiction as an \texttt{ERRATUM}. This separates two forms of success that are often conflated: \emph{a claim can fail while the verification process succeeds}.

That distinction also prevents a superficially perfect score from obscuring the result. The paper reports 95/96 testable claim criteria met and 148/148 verifier checks passing because the latter includes correct detection and classification of the failed Fisher claim. A reviewer can therefore audit not only the positive claims but the process by which a negative result remains visible.

\subsection{What is genuinely stronger than the first experiment}

The most important improvement is not that a larger number of tests passes. It is that the later evaluation can distinguish useful operation from refusal.

At v0.1, a verifier that rejects every output looks superficially ``safe'' but provides no evidence of selectivity. V2 introduces clean controls, severe injections, strictness sweeps, policy ablations, independent runtime witnesses, and adversaries that are guaranteed to clear the format boundary. Those additions make positive and negative outcomes simultaneously observable.

V3 then broadens the object of evaluation. A policy boundary is only as trustworthy as its provenance, failover, certificate verification, transaction semantics, and replay. The full-stack tests show that these pieces are at least coherently implemented under the declared synthetic model, while the claim register prevents unsupported pieces from borrowing credibility from the pieces that pass.

\subsection{Why the synthetic adversaries matter, and why they are not enough}

The confident-hallucinator surrogate is a stronger containment test than malformed Qwen output because it deliberately reaches the admission gate and asks for an action that should not always be permitted. The evidence-forger is stronger in a different direction because it supplies a plausible action while corrupting evidence provenance. Zero silent promotions under both shows that distinct gates are load-bearing.

However, neither adversary adapts to the full implementation, discovers unknown policy gaps, exploits side channels, or produces naturalistic model behavior. A frontier-model study should therefore follow, but it should preserve the v2 separation between interface admission and task-level correctness so that future models cannot be mischaracterized by formatting artifacts.

\subsection{The next decisive experiments}

The next experiments should retire the largest remaining uncertainties, in roughly this order:

\begin{enumerate}
  \item \textbf{Frontier-model evaluation under the non-degenerate v2 contract.} Run multiple model families with structured-output interfaces and preserve separate admission/task/policy metrics.
  \item \textbf{Actual BeTaL LLM designer.} Replace the deterministic feedback controller with the intended LLM-in-the-loop designer and evaluate cost, convergence, transfer, and reachability-aware behavior.
  \item \textbf{Human policy elicitation.} Have domain experts define severe classes, policy precedence, and adjudication denominators before seeing system outcomes.
  \item \textbf{Real systems integration.} Stage a FHIR server, attestation-capable TEE, and BFT implementation; measure the currently proxied or out-of-scope latency/failover claims.
  \item \textbf{Prospective or shadow-mode validation.} Only after the earlier layers are stable should real operational false-conflict and sensitivity claims be attempted.
\end{enumerate}

These experiments would transform the current result from a synthetic control-plane validation into evidence about deployed evaluation and governance.

\section{Conclusion}

The strongest result in this study is not a perfect pass rate. It is that the evaluation stack can represent different kinds of failure without converting them into the same flattering score. A model can fail before semantic scoring; a verifier can appear perfect by refusing everything; a benchmark optimizer can chase an unidentifiable or unreachable target; and a research paper can carry a numerical claim that its own stronger validation criterion disproves. In the last case, the correct system behavior is to preserve the erratum, not to make the dashboard green.

BeTaL-GBI v0.2 adds an admissibility-aware tuning objective and demonstrates model-free dynamic benchmark search over a 2.2-billion-point design grid. GBI v2 replaces answer-key runtime decisions with witness-grounded policy, restores non-vacuous selectivity, catches all declared severe injections under its selected policy, exposes denominator sensitivity, and blocks two distinct synthetic adversaries without silent promotion. GBI-DCSE v3 converts the full architecture into a 99-claim executable register, meets the declared acceptance criteria for 95 of 96 testable registered claims, leaves three claims explicitly out of scope, discovers one numerical erratum, and passes 148 declared verification checks. It further demonstrates signed-ledger behavior, attestation fail-closed logic, a tight modeled BFT fault boundary, a load-bearing sparse certificate, atomic rollback, receipt replay, and code-level portability to a second synthetic domain.

The defensible result is therefore not that GBI-DCSE has proved enterprise safety. It is that the architecture has progressed from a conceptual and fail-closed boundary to a substantially more testable, selective, versioned, and self-auditing verification substrate. The remaining work is also clearer: run real frontier models against the repaired boundary, use the actual LLM benchmark designer, elicit policy from domain authorities, and validate the systems layer in real deployment environments.

\appendix

\section{Full BeTaL-GBI parameter grid cardinality}

The grid cardinality is the product
\[
21\times 11\times 11\times 21\times 6\times 21\times 11\times 5\times 6
=2,218,750,380.
\]
The dimensions correspond respectively to patient-identity normalization, orphan rate, field-anomaly bleed, code/version validation, mapping arity, temporal ambiguity, evidence sufficiency, policy conflict depth, and distractor actions.

The search does not enumerate this grid. Phase 1 follows a monotone dial and Phase 2 performs local coordinate refinement. Accordingly, a target that appears unreachable on the dial may still be reached off the path, and a local optimum is not a global optimum claim.

\section{GBI v2 gate interpretation}

The declared policy is deliberately external-validity oriented. A decision is a function of declared witness evidence, policy version, validity, dependencies, identity state, and provenance. Benchmark reference actions remain useful for evaluation but are not a deployment primitive.

For injected records, the original reference action may become stale by construction. This explains why all 396 non-injected records can agree with their references while aggregate agreement is lower. A correct post-injection refusal can disagree with the pre-injection reference without being an implementation error. The paper therefore reports witness-grounded verdicts and reference agreement separately.
\section{V3 out-of-scope claims}

Three claims remain explicitly out of scope:

\small
\begin{longtable}{>{\RaggedRight\arraybackslash}p{0.13\linewidth} >{\RaggedRight\arraybackslash}p{0.38\linewidth} >{\RaggedRight\arraybackslash}p{0.39\linewidth}}
\toprule
Claim & Why out of scope & What is established instead\\
\midrule
C-9.16 & No zero-knowledge proof system is present. & Finite certificate predicate, public-input binding, and a blinded binding commitment.\\[0.6em]  \vspace{-0.4em} \\
C-B1.7 & Retrospective clinical playback requires a preregistered cohort and expert adjudication. & Receipt-driven replay reproduces 14/14 synthetic decisions.\\[0.6em]  \vspace{-0.4em} \\
C-B2.4 & Attestation bootstrapping target requires real TEE quote generation and an attestation-service handshake. & Software verification path is measured only as a proxy and is not promoted to the hardware target.\\
\bottomrule
\end{longtable}

\end{document}